\input{aipcheck}

\documentclass
  {aipproc}

\usepackage{amsmath}
\usepackage{amsfonts}
\usepackage{amssymb}
\layoutstyle{6x9}

\begin{document}

\title{Resonance-like Goos-H\"anchen Shift induced by nano-metal films}


\author{R. Gruschinski$^+$, G. Nimtz $^*$, and A.A. Stahlhofen$^+$}
{address={ *II. Physikalisches Institut, Universit\"at zu K\"oln and $^+$Institut f\"ur Integrierte Naturwissenschaften, Universit\"at
Koblenz}
}

\begin{abstract}
The influence of nano-metal films  on the Goos-H\"anchen shift (GHS) is investigated. The films deposited at the total reflecting surface of a perspex prism/air have a
sheet resistance varying between Z$_{\Box}$ = 25 and 3 000 $\Omega_{\Box}$.
A resonance-like enhancement of the shift and of the absorption is found for TE polarized waves,
when the sheet resistance approaches the value of the vacuum impedance. For TM waves the influence of the metal films on the GHS
is comparatively weak. The experiments are carried out with microwaves.
\end{abstract}

\maketitle

\section{Introduction}
A shift of the reflected electromagnetic wave against the incident beam in the case of total reflection was conjectured by Newton 300 years ago. Goos and H\"anchen have measured the shift around 1946 \cite{Goos} for the first time.
A quantitative experimental analysis of the shift was performed by Haibel  et al. with microwaves~\cite{Haibel}. In the study the shift was measured depending on the wave polarization, on the angle of incidence, and on the beam diameter. The result was that the reflected TM waves are much more shifted than the TE waves. The GHS was found to be of the order of magnitude of the wavelength. In addition
the shift was observed to increase with decreasing beam diameter.

For the time being there is no exact theoretical description of the GHS available. Moreover, the theoretical and some experimental
investigations predict a discontinuity of the GHS at the critical angle and no shift at angles below the critical one, see for instance Refs.\cite{Artman}. Recently, however, it was shown by M\"uller et al. \cite{Mueller} that there is a continuous transition from the angles of total reflection to angles of partial reflection. The novel result is, also partial reflection shows a significant beam shift between the incident and the reflected beam.

In this study
we present data of the total reflection influenced by nano metal films deposited on the total reflecting surface. The Al-films were between 10 and 100 $nm$ thick and were vapor deposited on polyethylene films of 10 $\mu m$ thickness. In addition we have used conducting films with higher sheet resistance values ($\ge$ 1000 $\Omega$)  based on an organic metal (\cite{Ormecon}). The organic material having a higher resistivity was sprayed also on polyethylene films of 10 $\mu m$ thickness. We applied single films as well as stacks of several sheet layers. The latter behaved as a parallel circuit with a corresponding lower total sheet resistance.

Incidentally, Woltersdorff ~\cite{Woltersdorff} has shown for thin metal films at low frequencies that the absorption A has a maximum with
A = 0.5 and T = R = 0.25 hold at the thickness $d_W$

\begin{eqnarray}
d_W = \frac{2}{Z_0 \sigma},
\label{GHS}
\end{eqnarray}

where $Z_0$ = 377 $\Omega$ and $\sigma$  are the vacuum impedance and the conductivity, respectively. The measured sheet resistance $Z_\Box$ is given by

\begin{eqnarray}
Z_\Box = 1/(d \,\,\sigma) \\
d_W \sigma = \frac{2}{Z_0 }.
\label{GHS}
\end{eqnarray}

The thickness of the Al-films produced by vapor deposition on 10$\mu m$ polyethylene films could not be measured. It was estimated to be between 10 and 100 $nm$. Only the sheet resistance  $Z_\Box$ was precisely measured. It is assumed that the conductivity of the nano films is one to two orders of magnitude smaller than the bulk conductivity \cite{Dumpich}. In addition to $Z_\Box$ we have measured the reflection R and the transmission T of the microwave at normal incidence.
The following relations are valid for the low frequency case

\begin{eqnarray}
R + T + A & = & 1 \,\,\, with\\
R & = & (1 + \frac{2}{\sigma d Z_0})^{-2} = (1 + \frac{2 Z_\Box }{Z_0})^{-2}\\
T & = & (1 + \frac{\sigma d Z_0}{2})^{-2} = (1 + \frac{Z_0}{2 Z_\Box })^{-2},
\end{eqnarray}

\begin{figure}[htb]
\includegraphics[width=1\textwidth]{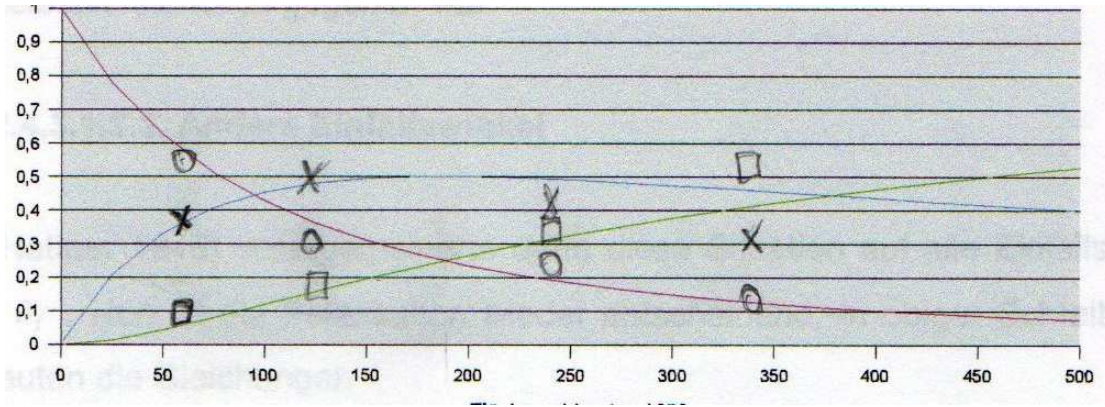}
\caption{A,R,and T vs sheet resistance.
\label{ART}}
\end{figure}

The relations with experimental values are presented in Fig.\ref{ART}. The following study was carried out with  Al and metallic organic films near the absorption maximum

\begin{figure}[htb]
\includegraphics[width=0.7\textwidth]{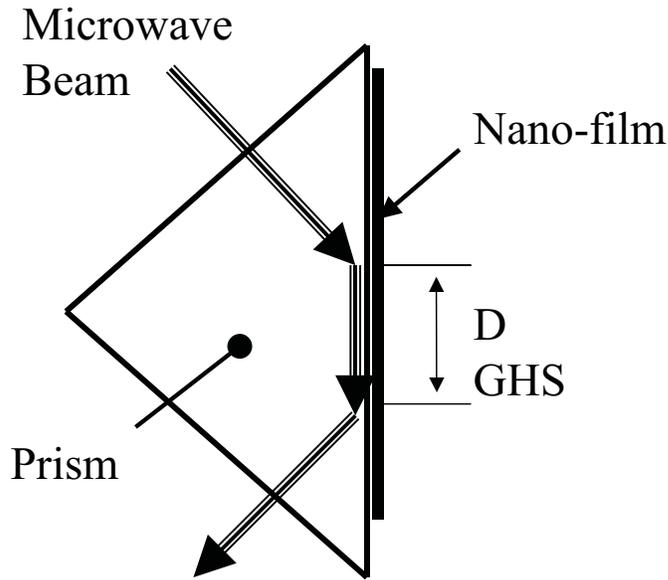}
\caption{Sketch of the experimental set-up. The perspex prism. A parabolic antenna transmitted a beam of microwave pulses of 8 ns half width. Details are the same as in Ref.\cite{Haibel}. The receiver was a horn antenna and a fast rectifier.
\label{set-up}}
\end{figure}

\begin{figure}[htb]
\includegraphics[width=0.9\textwidth]{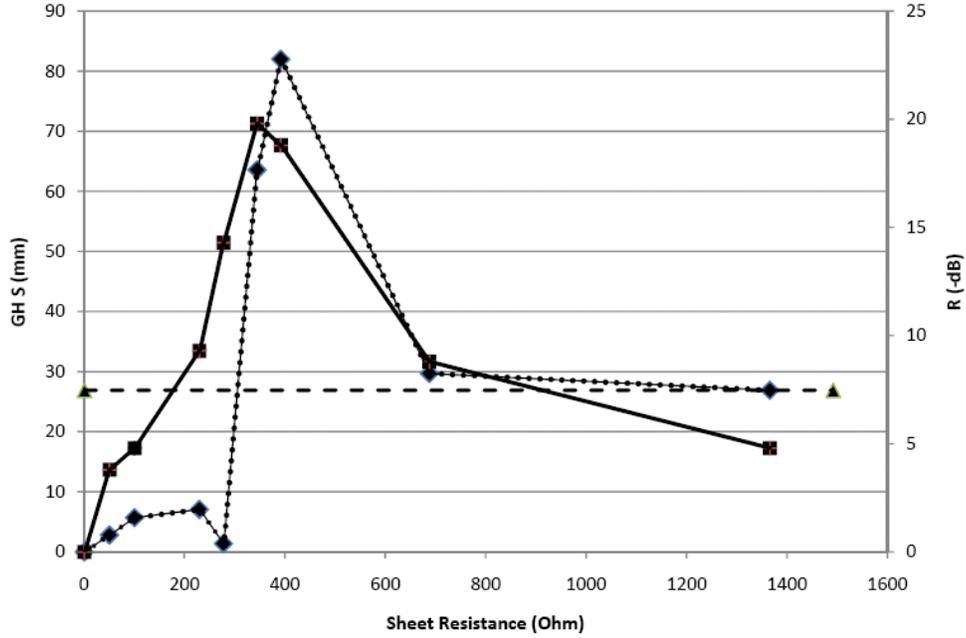}
\caption{Experimental data of GHS and reflection vs sheet resistance for TE waves. The broken line represents the total reflected GHS of 26.9 $mm$ without a metal film on the reflecting surface. The GHS increased up to 82 $mm$ and the reflection decreased up to -20 $dB$ at a sheet resistance near 377 $\Omega$. The values at zero resistance correspond to the data of a bulk aluminum plate at the reflecting surface. The errors are $\pm$  1 $dB$ and $\pm$ 5 $mm$.
\label{GHS2}}
\end{figure}

\begin{figure}[htb]
\includegraphics[width=0.9\textwidth]{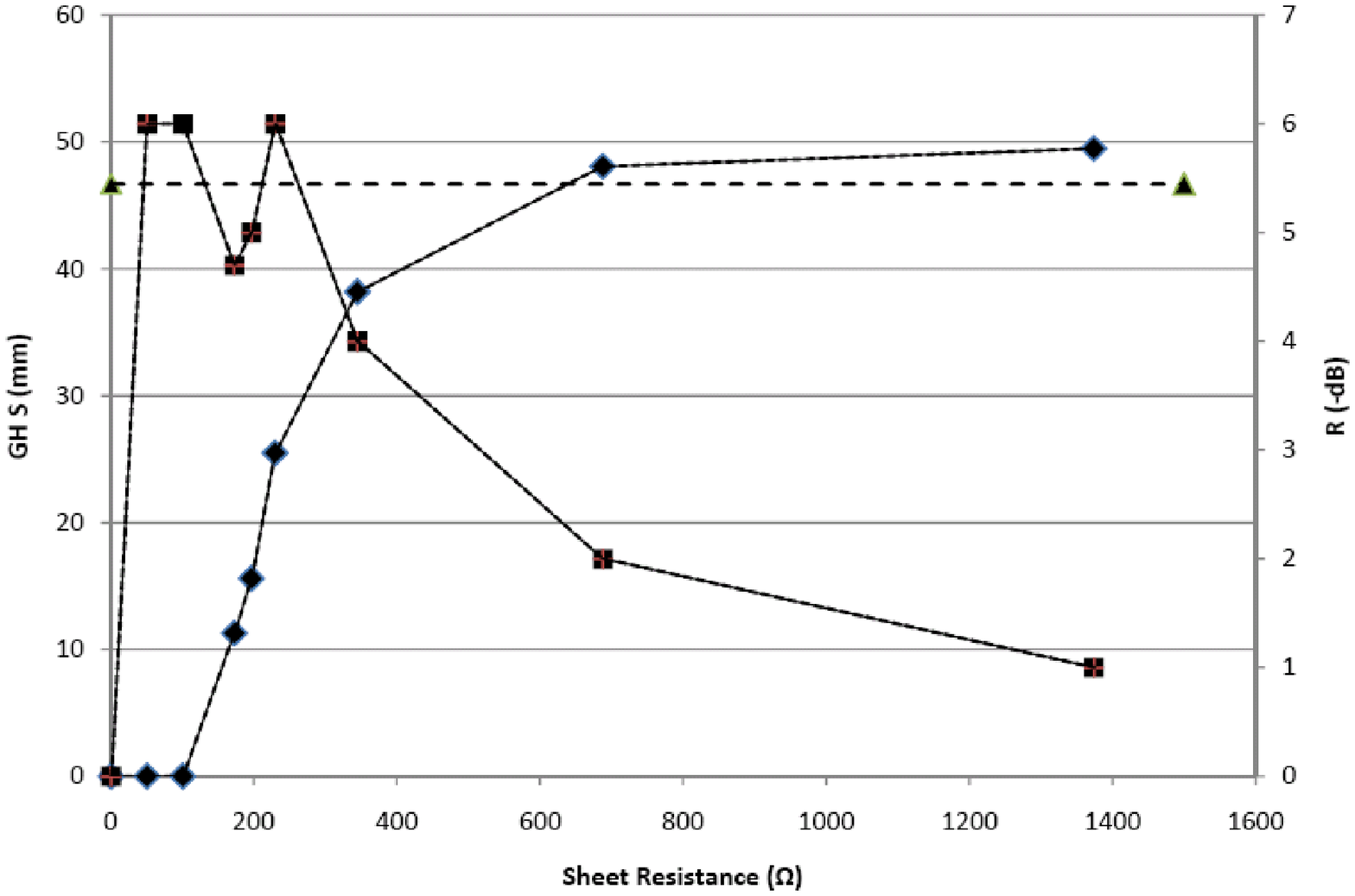}
\caption{Reflection and GHS vs sheet resistance $Z_\Box$ for TM polarized waves.
The data at zero sheet resistance corresponds to a bulk aluminum mirror. The dotted line represents the GHS of 46.7 $mm$ for the prism-air configuration. The errors are
$\pm$  1 $dB$ and $\pm$ 5 $mm$.
\label{GHS1}}
\end{figure}

\section{Experiment} The total reflection was investigated at a perspex prism of 400 $mm$ height and side length. The refractive index was 1.605 corresponding to a critical angle of $38.5^o$. The wavelength and the frequency of the microwaves are 32.8 $mm$ and 9.15 $GHz$. The parabolic antenna transmitted parallel waves. The beam width was limited by an aperture of 120 $mm$ diameter. The angle of incidence was $45^o$. As shown in the sketch of Fig.\ref{set-up} the metal films were placed on the reflecting surface. The metal films were produced by aluminum evaporated on 10 $\mu m$ polyethylene substrates. Alternatively we used organic conducting poly-aniline films sprayed also on 10 $\mu m$ polyethylene substrates~\cite{Ormecon}. Layer thickness including substrate of all the applied films was $\ll$ than $\lambda$. The aluminum films are about 20 $nm$ thick. The sheet resistances were between 50 and 2746 $\Omega_\Box$.

In the case of TE polarized waves the reflection decreases with increasing impedance with a minimum of 20 $dB$ near 377 $\Omega_\Box$, however,
at higher sheet resitances the reflection increases approaching the GHS and the mirror value finally as shown in Fig.\ref{GHS2}. At the reflection minimum (i.e. the maximum of signal amplification in Fig.\ref{GHS2}) the power is absorbed and not transmitted. More important is that at the reflection minimum the GHS has a strong maximum: three times the normal GHS.

Experimental data of the reflection and the GHS  for TM polarized waves is plotted in Fig.\ref{GHS1}. In turn the zero impedance values are the values measured with a bulk  aluminum plate as reference, the GHS without films is also presented. At small resistances a GHS is not observed, with increasing resistance of the metal films the data approaches the GHS value.

\section{Discussion}
There is no exact theory of the GHS available for the time being. Besides the quantitative deviations between experimental and theoretical data two main problems are
not solved: The GHS diverges to large values when approaching the critical angle and the observed dependence on the beam diameter. As mentioned above there is a continuous transition from total to partial reflection as is shown in Ref.\cite{Mueller}. There is one agreement of the experimental results and the theoretical data since Artman's study: the GHS is larger for TM than for TE polarized waves. The relationship for the GHS D of the reflected beam is given by the proportionality

\begin{eqnarray}
D  & \propto &  - \frac{1}{k_1} \,\,\,\, \frac{d\phi}{d\theta_{in}}, \label{E}\\
D_{TM} & = &  (\frac{n_1}{n_2})^2 \,\, D_{TE} ,
\label{GHS}
\end{eqnarray}

where $k_1$ is the wave number of the incident wave, $n_1$ and $n_2$ are the refractive indices of the first and the second optical media with $n_1$ $>$ $n_2$. $\theta_{in}$ and $\phi$ are the angles of incidence and $\phi$ the phase shift between the incident and the reflected wave.

It is interesting that the anomalous behavior of R and GHS
 for TE polarization happens when the metal film layer has a resistance equal to the vacuum (air) impedance.
 Obviously, at this impedance the thin metal film matches the air medium resulting in a long surface beat mode with a high absorption of the TE mode. The electric field is now in the plane of the conducting layer. In the case of TM polarization the electric field is perpendicularly orientated to the metal film and is less phase shifted and less absorbed.

The resonance like anomaly can not be related to a surface plasmon resonance because this plasmon enhancement is only expected and observed for transverse magnetic excitation e.g. Ref.\cite{Yin}.

The conductivities $\sigma$ of the films result in a low frequency complex refractive index, where both components are of equal value

\begin{eqnarray}
n & = & n_1 - i\,\, n_2 \\
n_1 & \approx & n_2 \approx \sqrt{\frac{\sigma}{4 \pi \epsilon_0 \nu}} \,\,\,\,\,with\\
100  \geq n_1 & \approx & n_2 \leq 1000,
\end{eqnarray}
where $\epsilon_0$ is the vacuum permittivity and $\nu$ the electromagnetic wave frequency. This extremely strong step of the refractive index makes the observed
maximum of the GHS for TE polarized waves plausible, having in mind Eq.\ref{E} and $d\phi$ $\propto$ $dn$. A quantitative theoretical approach is not available  yet.

\section{Summary}
Experiments on total reflection and GHS under the influence of nano-metal films with a thickness $d$ $\ll$ $\lambda$ the wavelength and near
the Woltersdorff thickness $d_W$  show a resonance like  behavior for the TE polarized wave near the vacuum impedance. The reflection has a strong minimum of -20 dB compared to total reflection or reflection by a bulk aluminum mirror.
At the same time the GHS shows a large peak triple the value of the normal GHS. The TM polarized wave is not much effected since the
electric field is normal to the metal films.

\begin{theacknowledgments} We gratefully acknowledge the technical support by U. Blancke, A. Enders, and U. Panten.

\end{theacknowledgments}

\end{document}